\documentclass[12pt]{book}

\usepackage[dvips]{graphicx,color}
\usepackage{makeidx,view}

\makeauthorindex
\makeindex

\begin{document}

\BookTitle{\itshape The Universe Viewed in Gamma-Rays}
\CopyRight{\copyright 2002 by Universal Academy Press, Inc.}
\pagenumbering{arabic}

\chapter{Diffuse Emission from the Galactic Plane and Unidentified EGRET Sources}%

\author{%
Martin Pohl\\
{\it Institut f\"ur Theoretische Physik, Lehrstuhl 4, Ruhr-Universit\"at Bochum, 44780 Bochum,
Germany}}

%
\AuthorContents{M.\ Pohl} 

\AuthorIndex{Pohl}{M.}

\section*{Abstract}
Diffuse Galactic gamma-ray emission is produced in interactions of
cosmic rays with gas and ambient photon fields and thus provides us
with an indirect measurement of cosmic rays in various locations
in the Galaxy. The diffuse gamma-ray continuum is more intense than
expected both at energies below 200 keV and above 1 GeV. The existing
models for the high-energy excess are reviewed in the light of recent
TeV gamma-ray measurements of both diffuse emission and of discrete
Galactic sources such as supernova remnants. I specifically discuss whether
particular classes of Galactic objects are observable as
EGRET gamma-ray sources, either by being found among the
unidentified EGRET sources or by contributing to the diffuse emission as
unresolved sources.

\section{Introduction}
The EGRET instrument aboard the Compton Gamma-Ray Observatory, CGRO, has observed the 
Galactic diffuse gamma-ray emission with unprecedented detail and accuracy. This emission
is supposedly produced in interactions of Galactic cosmic rays, electrons and protons,
with the interstellar gas and the interstellar radiation field. The spectral and spatial
distribution of the diffuse radiation can be compared with models based on the locally
observed spectra of cosmic rays and the Galactic distribution of interstellar gas
and soft photon fields. Such studies indicate that while at photon energies below 1 GeV
the observed intensity distribution in the Galactic plane
is in accord with the model predictions, at higher energies above 1 GeV the observed
intensity in the Galactic plane displays a GeV excess at a level of 60$\%$ 
compared with the predictions (Hunter et al. 1997). The diffuse Galactic
emission around 100 keV is also in excess of predictions based on cosmic
ray propagation models (Purcell et al. 1996; Valinia, Kinzer, and Marshall 2000).

EGRET has detected about 270 point sources, about two thirds of which have not
been identified with objects observed in other frequency regimes 
(Hartman et al. 1999). It is possible, that the gamma-ray excesses are related
to the unidentified point sources.
This review will address the four following questions:

\noindent
$\bullet\ $What is the nature of the excesses in diffuse Galactic gamma-rays?

\noindent
$\bullet\ $What fraction of the diffuse Galactic emission is caused
by unresolved sources?

\noindent
$\bullet\ $What is the nature of the unidentified EGRET sources?

\noindent
$\bullet\ $What is the relation between the gamma-ray excesses and the
unidentified gamma-ray sources?

\section{What is the nature of the excesses in diffuse Galactic gamma-rays?}

The diffuse Galactic gamma-ray emission is most intense in the direction of the 
inner Galaxy. Figure 1 shows the observed intensity spectrum from that region.
\begin{figure}[t]
  \begin{center}
    \includegraphics[width=13.truecm]{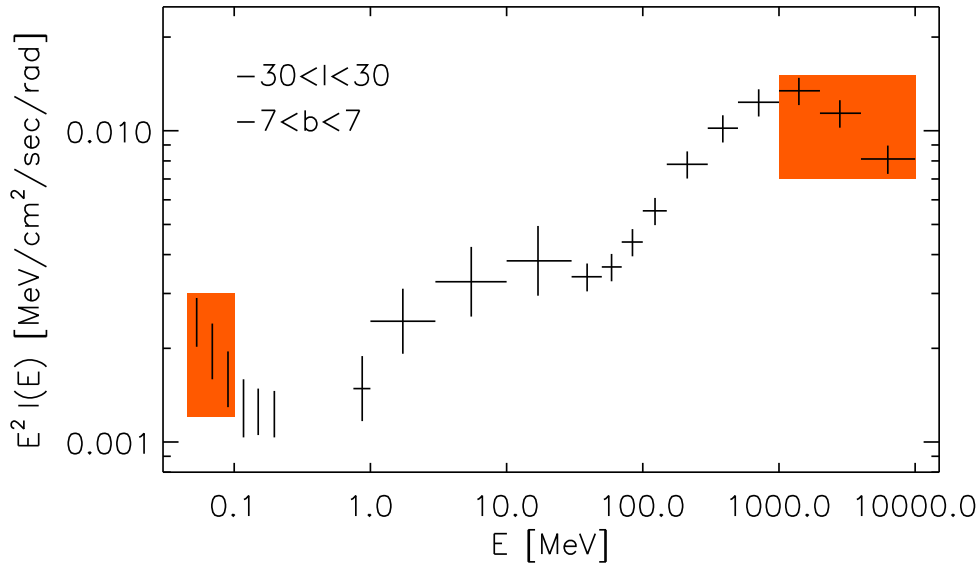}
  \end{center}
  \caption{The intensity spectrum of diffuse emission from the inner Galaxy.
The data around 100 keV have been obtained with the OSSE experiment 
(Kinzer, Purcell, and Kurfess 1999),
those in the MeV range result from observations with COMPTEL (Strong et al.
1996),
and the GeV range spectrum has been derived from EGRET data after subtraction of
known point sources (Hartman et al. 1999). The shaded regions indicate,
at which gamma-ray energies the observed intensity exceeds that expected
from interactions of cosmic rays with spectra similar to those observed in
the solar vicinity.}
\end{figure}
The shaded areas indicate, at which gamma-ray energies the observed intensity
exceeds that expected from interactions of cosmic rays with spectra similar to
those observed in the solar vicinity. The excess near 100 keV can, if 
truely diffuse, only be caused by either electron or proton bremsstrahlung.
In both cases the radiation efficiency would be very low, and therefore
a very high cosmic-ray source power in excess of, e.g., that supplied by
supernovae would be needed to sustain the particle population (Skibo et al. 1996; Valinia \& Marshall 1998).
In addition, proton bremsstrahlung as the main emission
mechanism of the observed low-energy gamma rays seems to be in conflict with
the observational limits on nuclear line and pion-decay continuum emission
(Pohl 1998). The point source contamination of the Galactic 100 keV
radiation is unknown, but will be determined with INTEGRAL in the near future.
\begin{figure}[t]
  \begin{center}
    \includegraphics[width=13.truecm]{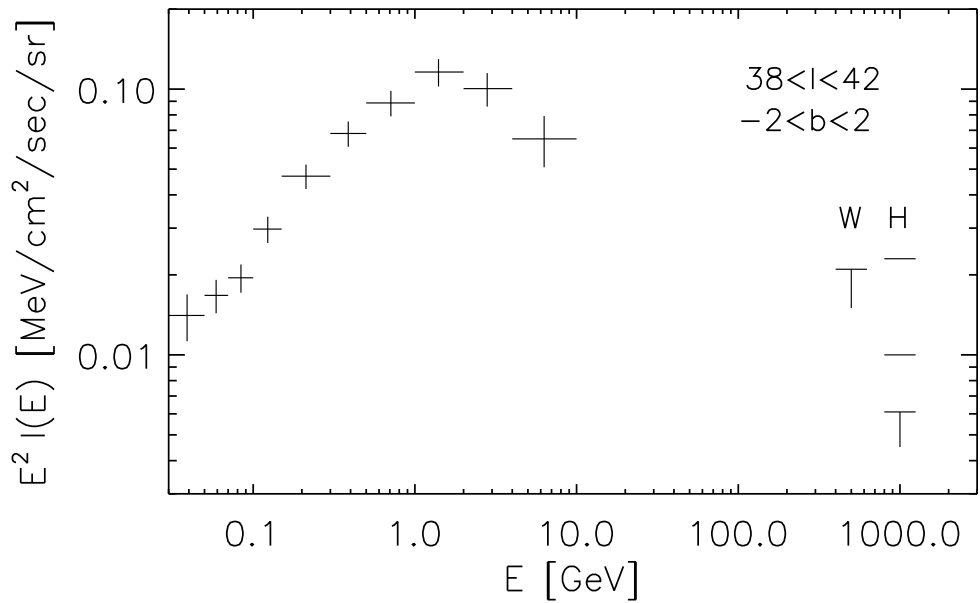}
  \end{center}
  \caption{The intensity spectrum of diffuse gamma-ray emission from a small region in
  the Galactic plane near $l=40^\circ$. The GeV-scale data have been obtained with
  the EGRET experiment. In the TeV range upper limits have been published by
  the Whipple team ("W", LeBohec et al. 2000) and the HEGRA collaboration
  ("H", Aharonian et al. 2001), for which three limits are given depending on whether 
  all gamma-ray like events are considered, or a high-latitude observation or 
the $\vert b\vert \ge 2^\circ$ data are used for background estimation.}
\end{figure}
\begin{table}[t]
    \caption{Results of a linear correlation analysis of the observed intensity
    versus that expected for the diffuse Galactic emission at high latitudes
    (Sreekumar et al. 1998). The residual, A, is the intensity in units of
    (${\rm 10^{-6}\ cm^{-2}\,sec^{-1}\,
    sr^{-1}}$) of the presumably extragalactic emission that remains after
    extrapolating to zero Galactic radiation,
    whereas the slope, B, is the scale factor needed to match model and
    observation of the Galactic emission. The large values of B for energies beyond 
    1 GeV indicate a GeV excess similar to that in the Galactic plane.} 
\begin{center}
\begin{tabular}{|c|c|c|} \hline
Energy range in MeV& Residual A & Slope B    \\   \hline
30-50      & 24.0$\pm$6.4   & 1.14 \\ 
50-70      & 13.1$\pm$1.9   & 1.04  \\  
70-100     & 7.8$\pm$0.27   & 1.09  \\   
100-150    & 5.5$\pm$0.20   & 1.05  \\  
150-300    & 5.3$\pm$0.21   & 0.97  \\  
300-500    & 1.9$\pm$0.10   & 0.97  \\  
500-1000   & 1.3$\pm$0.07   & 1.09  \\  
1000-2000  & 0.67$\pm$0.036   & 1.34  \\  
2000-4000  & 0.41$\pm$0.028   & 1.85  \\  
4000-10000 & 0.19$\pm$0.017   & 1.56  \\   \hline
\end{tabular}
\end{center}
\end{table}

The GeV excess is unlikely to be an instrumental effect, in particular because
the Crab spectrum is observed to be a single power-law with no
indication of an excess at GeV energies. The extent of the GeV excess towards
higher energies is not well constrained by the existing limits on diffuse TeV-scale emission, which are shown in Figure 2. The
EGRET data displayed in that figure are not corrected for the effects of the
point-spread function, so that the true intensity at energies below around 300
MeV is probably somewhat higher than indicated here. An extrapolation of the
spectrum at a GeV and higher, at which energies the point-spread function is a minor
concern, is not in conflict with the upper limits in the TeV range. The TeV data
do, however, constrain models of the GeV excess which invoke a hard spectral
component arising, e.g., from inverse Compton (IC) scattering of high-energy electrons
accelerated in supernova remnants (Pohl \& Esposito 1998).
The available X-ray
data from young remnants indicate that electrons are accelerated to around 10 TeV,
beyond which the spectrums shows a strong depression or cut-off (Reynolds
\& Keohane 1999), implying corresponding turn-overs in the IC spectra
around 100 GeV, so that the present TeV scale data are not in conflict with the 
hard IC interpretation of the GeV excess.

The original analysis that led to the detection of the GeV excess
concentrated on the Galactic plane, for the Galactic emission is strongest
in that region. 
One should note that also at high latitudes a corresponding excess
is visible. Sreekumar et al. (1998) have linearly correlated the observed intensity
with that predicted by a cosmic-ray propagation model with a view to determine
the extragalactic background by extrapolating to zero Galactic emission. 
Table 1 gives the slopes of that correlation at different energies. 
They are around unity for gamma-ray energies below
1 GeV, but vary between 1.35 and 1.85 above 1 GeV, indicating the existence of a GeV
excess of roughly the same magnitude as in the Galactic plane. 

\section{Unresolved sources in the diffuse gamma-ray emission}
Estimates of the contribution of unresolved sources have a firm basis only
if the flux level of at least a few members of the source class in question
is known. In the gamma-ray regime the only class of Galactic sources
with that property is pulsars. Emission models or just the luminosities
of the observed sources can then be used to estimate the gamma-ray intensity
produced by the total population of sources.
\begin{figure}[t]
  \begin{center}
    \includegraphics[width=13.truecm]{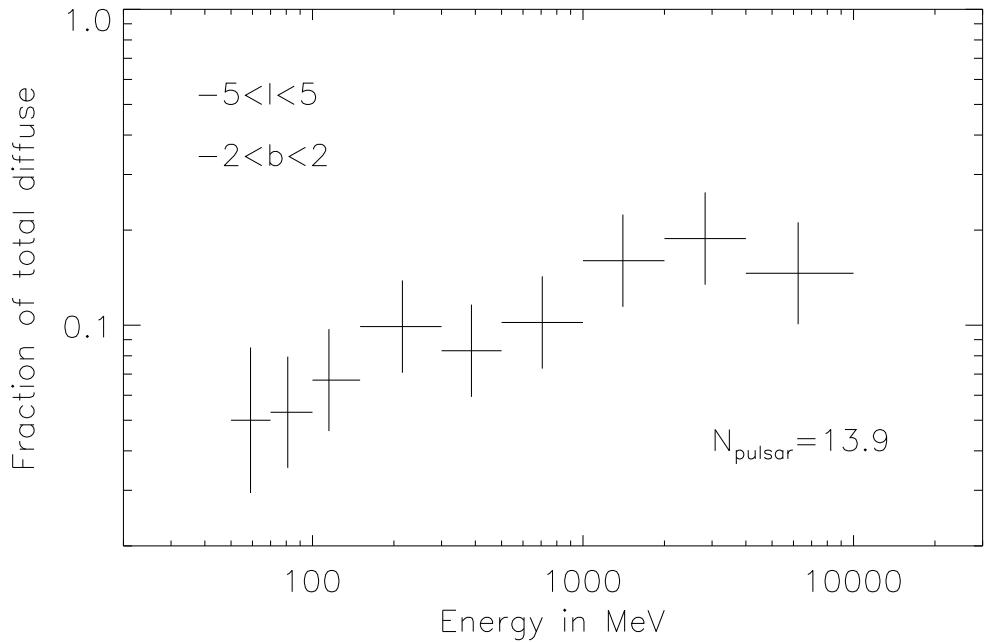}
  \end{center}
  \caption{The intensity spectrum of unresolved pulsars shown in units of
  the total observed diffuse gamma-ray emission (Pohl et al. 1997). The
  intensity scales linearly with the number of observable pulsars,
  which would be 13.9 for this plot.}
\end{figure}
Based on the observed spectra of six EGRET pulsars
Pohl et al. (1997) have fitted an empirical luminosity function at
different energies and were thus able to estimate the intensity spectrum
produced by unresolved pulsars. The absolute intensity levels scales
linearly with the number of pulsar observable as point sources
(not necessarily as pulsars). Figure 3 shows the fraction of the total diffuse
emission from the Galactic center region that would be produced by unresolved
pulsars. Pulsars would indeed most significantly contribute at energies of 
a few GeV. In the Galactic center direction, where the pulsar
contribution would be strongest, they would provide nearly 20\% 
of the observed intensity above 1 GeV for 14 observable pulsars. In 
other directions the contribution would be less than that. Figure 4
shows the latitude distribution of the intensity of unresolved pulsars
averaged over Galactic longitude. To be noted from the figure is that 
on average the pulsar contribution to the diffuse Galactic emission is below
10\% and that the predicted latitude distribution differs from that of the
observed diffuse emission, so that at higher latitudes the pulsar contribution
becomes negligible. Thus unresolved pulsars cannot explain the GeV excess.
\begin{figure}[t]
  \begin{center}
    \includegraphics[width=13.truecm]{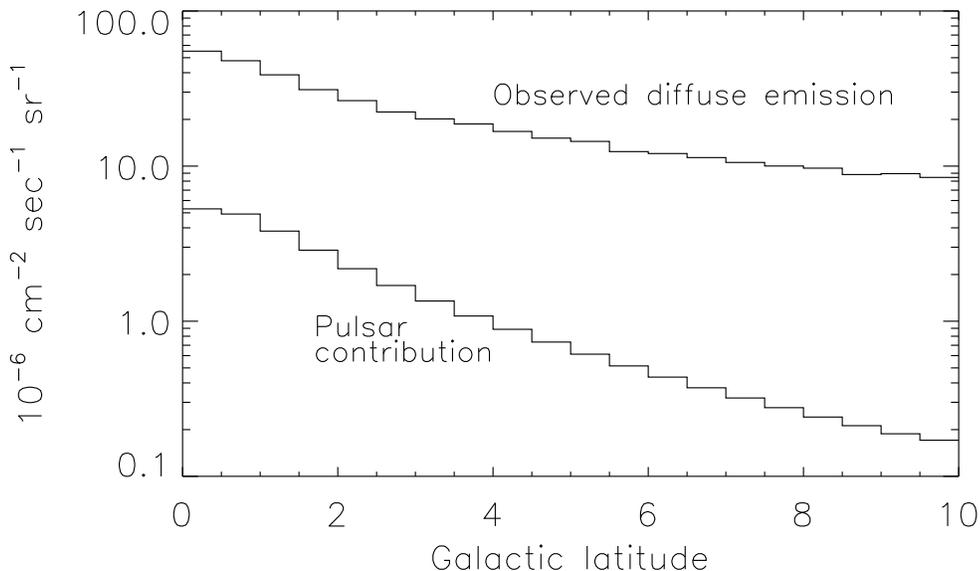}
  \end{center}
  \caption{The longitude-averaged latitude distribution of the intensity
  provided by unresolved pulsars shown in comparison with 
  that of the observed diffuse emission. The absolute
  intensity corresponds to 13.9 observable pulsars as in Fig.3.}
\end{figure}
McLaughlin and Cordes (2000) have used a spin-down model to estimate the
pulsar contribution to the diffuse Galactic emission above 100 MeV and the
number of pulsars among the EGRET unidentified sources. They find that about twenty unidentified sources might be pulsars. Nevertheless, pulsars would not
significantly contribute to the diffuse Galactic gamma-ray emission. These
results are completely in line with those of Pohl et al. (1997), though 
the methods used are different. 

\section{The nature of unidentified EGRET sources}
In this review we are mainly interested in Galactic sources. So how do we
know a source is galactic, if it is unidentified? Of course we can't tell
for the individual source. Intriguing though its association with Cygnus OB2 
is, the one and only unidentified TeV gamma-ray source could be a background
object, provided the indications of extended emission prove unsubstantiated
(Aharonian et al. 2002). 
If many unidentified sources are present, like 170 in case of EGRET, one can
use their
spatial distribution to infer how many of them have a Galactic
origin. Approximately 100 EGRET sources appear to be Galactic on this ground.
This result depends somewhat on the assumed spatial distribution
and on the luminosity function of the sources in question. Consequently
the uncertainty in that figure is slightly higher than the Poissonian
error margin of $\pm 10$.

A many authors have attempted to infer the nature of the unidentified Galactic
sources by studying the spectra (Merck et al. 1996; Zhang \& Cheng 1998), the
variability properties (McLaughlin et al. 1996), or
spatial correlations with known source classes such as supernova remnants
(SNR) or OB 
associations (Sturner \& Dermer 1995; Kaaret \& Cottam 1996,;
Romero et al. 1999). However, care must be exercised 
in the analysis of EGRET data, for a number of systematic effects
influence the population studies. Flat-spectrum sources are more easily found
in regions of a high background intensity, i.e. in the Galactic plane.
The backward-folding of data in the EGRET standard analysis, as opposed to the
forward-folding of a sky model through the instrument response, can cause a 
"false" variability of sources in the Galactic plane. Correlation studies
with known classes of objects are hampered by the large number of objects both
in the source class in question and in the typical EGRET error box for a
Galactic plane source. Nevertheless, few of the low-latitude unidentified
source seem to have a hard gamma-ray spectrum with index $\gamma \le 2.0$.

It may be useful then to calculate what one might expect to see from 
different source classes. We have already discussed the case of pulsars, for 
which one would expect a constant flux and a rather wide latitude distribution.
Only a few of the unidentified EGRET sources display the hard spectrum
that appears typical of the identified pulsars. It is possible, though,
that one observes pulsars by their off-beam emission, which in the polar-cap
model would have a soft spectrum with low luminosity (Harding \& Zhang 2001).
Such off-beam gamma-ray pulsars would be candidates for the population
of unidentified EGRET sources that is apparently associated with Gould's Belt
(Grenier 2000; Gehrels et al. 2000). Off-beam emission is not expected in
outer-gap models (Yadigaroglu \& Romani 1995; Cheng \& Zhang 1998), which
predict that most of the unidentified EGRET sources in the Galactic plane are
radio-quiet pulsars.
\begin{figure}[t]
  \begin{center}
    \includegraphics[width=13.truecm]{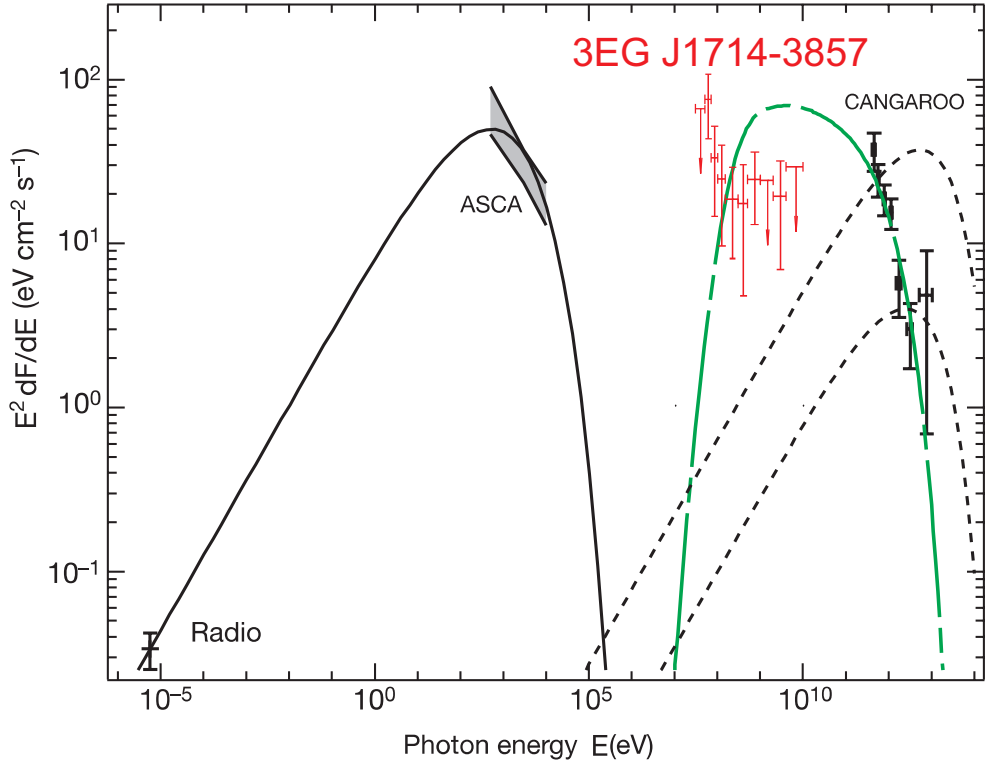}
  \end{center}
  \caption{The multi-band spectrum of RX J1713.7$-$3946 from 
Enomoto et al. (2002), revised by Reimer \& Pohl (2002)
to include the $\gamma$-ray spectrum of 
3EG~J1714$-$3857, is shown in comparison with emission
models presented in the former publication. The solid line indicates 
synchrotron emission, and the dashed lines are
the corresponding IC spectra based on the microwave
background and the ambient far-infrared photon field for two sets
of parameters, both of which would
violate the observed TeV $\gamma$-ray spectrum.
The green (grey) long-short-dashed line shows the $\pi^0$-decay spectrum,
which significantly exceeds the total observed emission by a factor of three.}
\end{figure}

SNR should also display a constant flux, but would
have a narrow latitude distribution. No EGRET source has been unambiguously 
identified with a SNR to date, but three SNR have been observed as sources of 
TeV gamma-rays. It is difficult to extrapolate the gamma-ray spectrum of SNR 
from the TeV band to GeV energies, for the relative importance of the 
contributions 
from $\pi^0$-decay, non-thermal bremsstrahlung and IC scattering
are a priori unknown. If a non-thermal spectral component exists at X-ray 
energies, which most likely would be synchrotron emission
of extremely high-energy electrons, the flux and the spectral form of the
IC emission depends only on the magnetic field
strength at the rim of the remnant and in its interior. The flux of gamma-rays
from $\pi^0$-decay and from bremsstrahlung cannot be easily estimated because
the density of both the non-thermal particles and the ambient gas is unknown.

It is generally difficult to infer the main radiation mechanism in the TeV band.
In the case of SN~1006 the dominant process is probably inverse Compton 
scattering on account of the low-density environment in which the remnant
resides. The situation is less clear for RX~J1713.7$-$3946 and Cas~A, for
which, however, there is also no clear evidence for hadronic TeV-scale emission.
A good example of the dilemma is RX~J1713.7$-$3946, whose spectrum is shown
in Fig.5.
Recent measurements with the CANGAROO II telescope have indicated that the
TeV-scale gamma-ray spectrum of RX~J1713.7$-$3946 can be well represented by a
single power-law with index $\alpha \simeq 2.8$ between 400~GeV and 8~TeV
(Enomoto et al. 2002). The authors argue that the multi-band spectrum from
radio frequencies to TeV gamma-ray energies cannot be explained as the
composite 
of a synchrotron and an IC component emitted by a population of 
relativistic electrons. It is then claimed that the spectrum of the
high-energy emission is a good match to that predicted by pion decay. Hence
RX~J1713.7$-$3946 would provide observational evidence that protons are 
accelerated in SNR to at least TeV energies.

Then Reimer \& Pohl (2002) have reanalyzed the multi-band spectrum of 
RX~J1713.7$-$3946 under the constraint that the GeV-scale emission
observed from the closely located EGRET source 3EG~J1714$-$3857 is taken
into account as either being associated with the SNR or as an upper limit
to the emission of the SNR. For both cases they find that a pion-decay
origin of the observed TeV-scale gamma-ray emission of RX~J1713.7$-$3946 is 
highly unlikely, contrary to the previous claim. The paucity of
spatially resolved radio and X-ray data suggests that answering the
question whether or not IC scattering can be responsible 
for the observed TeV-scale gamma-ray emission of RX~J1713.7$-$3946, whatever the
origin of the EGRET source 3EG~J1714$-$3857, requires a better knowledge
of the synchrotron spectrum of the SNR than available to date.

\section{The relation between the gamma-ray excesses and the
unidentified EGRET sources}
Whatever the dominant emission process of the TeV-scale emission from
Cas~A and RX~J1713.7$-$3946, the relative sensitivities of EGRET and
the present generation of atmospheric \v Cerenkov telescopes 
in conjunction with the expected form of the gamma-ray spectrum from SNR indicate that either only a few SNR can be among the EGRET unidentified
sources or many SNR do not accelerate cosmic rays to more than 10 TeV, in which
case the gamma-ray spectra would display a turn-over near 100 GeV.

But what about unresolved SNR, would they contribute to the diffuse
Galactic gamma-ray emission? The expected gamma-ray spectrum of unresolved
SNR would be substantially harder than that of the diffuse Galactic emission,
because the energy-dependent escape of cosmic rays from the Galaxy has no
effect inside the sources. Unresolved SNR would therefore most significantly
contribute in the TeV band (Berezhko \& V\"olk 2000), but presumably very
little at GeV energies. This is in accord with the energy-dependence
of the Boron-to-Carbon ratio observed in the solar vicinity. If cosmic rays
would predominantly interact with gas while still residing in their sources,
essentially all of the Boron production would also occur within the remnants,
and the Boron-to-Carbon ratio would therefore have to be flat, contrary to
the measured behaviour up to 20 GeV/nuc. Thus unresolved SNR contribute little
to the diffuse Galactic GeV-scale emission and in particular can not be the 
source of the GeV excess.

If unresolved SNR do not account for the GeV excess, what are the
consequences for the emission of electrons that have already been released
by the remnant? One of the possible explanations for the GeV excess
is the Swiss Cheese Model (Pohl \& Esposito 1998; Strong, Moskalenko,
and Reimer 2000), which argues that high-energy electrons suffer energy losses
so rapidly, that they do not propagate very far from their sources. 
Consequently the spatial distribution of these electrons would be inhomogeneous
and the locally observed spectrum would not be representative for the
electron spectra 
at other places in the Galaxy, where they could be substantially harder.
The correspondingly hard IC gamma-ray spectrum, so the idea, would explain
much of the GeV excess. 

In these calculations it was assumed that the electrons are instantly released
when accelerated, and that the electron acceleration would proceed for
$10^4$ to $10^5$ years in a typical remnant. GeV-scale IC emission is produced
by TeV-scale electrons, which have a radiative lifetime of around $10^5$ years. 
During that time the average electron will propagate about 300 pc, so that 
the SNR would be embedded in a cloud of high-energy electrons with
radius 300 pc. Seen from a distance, say 5 kpc away, the corresponding 
enhancement in IC intensity would have a radius of about $3^\circ$. The 
enhancements at distances less than 5 kpc from us would not be point sources
for EGRET and, consequently, would be subsumed with the diffuse Galactic 
emission as originally assumed.

But is that realistic? We know that cosmic rays are not instantly released
by the remnants, because the scattering mean free path of cosmic rays is 
much smaller near the rims of the remnant than it is in interstellar space.
After $10^5$ years the SNR has expanded to a radius of 50$-$100 pc, depending
on the density of gas in its environment, corresponding to approximately
$0.5^\circ -1^\circ$ when seen from 5 kpc away. Upstream of the (parallel) SNR shock,
in the precursor, the density of high-energy electrons falls off on a scale
$\delta r\simeq \kappa /V$, where $\kappa$ is the diffusion
coefficient and $V$ is the shock velocity. In the Sedov case $V\simeq 300\ 
$km/sec after $10^5$ years and $\kappa = \eta \kappa_{\rm B}$ will be
many orders of magnitude larger than the Bohm diffusion coefficient, 
$\kappa_{\rm B}$, i.e. $\eta \gg 1$. Then for TeV electrons in a
typical magnetic field of $B= 5\ \mu$G we obtain $\delta r \simeq \eta\, 0.1\
$pc$\,\gg\, 1\ $pc. We do not know the value of $\eta$, but it is probably very large,
so that after $10^5$ years the enhancement in the density of
high-energy electrons will have a radius at some value between the size of the
remnant, that is 50$-$100 pc, and the radius calculated under the
assumption of instantaneous release, approximately 300 pc.

What are the consequences? First, the corresponding enhancement in the
IC intensity would still have a radius of around $1^\circ$ when 5 kpc away,
so that nearby 
structures, e.g. those at Galactic latitudes of $5^\circ$ and higher,
would still not appear as point sources in the EGRET data. Second, the smaller
the region of enhanced electron flux around the location of a supernova is,
the larger are the temporal fluctuations in the electron flux at a
given location, which actually enhances the compatibility of the locally
observed high-energy electron spectrum with the electron
source spectrum that is required in the Swiss Cheese Model.
All in all, the non-detection of individual SNR in the EGRET data does not
seem to be in a strong conflict with that model of the GeV excess.

We have seen that only few of the Galactic unidentified sources will be pulsars
and SNR. That leaves us with still $\sim$100 unidentified sources.
Is the GeV excess perhaps caused by an unknown class of Galactic gamma-ray
emitters, which we also find among the EGRET unidentified sources?
Torres et al. (2001) have analyzed 40 low-latitude unidentified sources
which are not positionally coincident with any known class of potential
gamma-ray emitters. Surprisingly, many of them appear to be variable and on 
average these sources have a soft spectrum. What we
observe is apparently a new class of gamma-ray emitters.
Obviously, unresolved sources of that class cannot contribute to the
GeV excess on account of the soft spectrum. We can calculate a lower limit
for their luminosity $L\ge 10^{34}\ $erg/sec, though, by arguing that
they should not overproduce the diffuse emission around 100 MeV.
 
\section{Summary}
In this paper I have discussed the relation between unidentified EGRET sources
and the diffuse Galactic gamma-ray emission with emphasis on the GeV excess.
The results can be summarized as follows:

\noindent
$\bullet\ $About 100 of the EGRET unidentified sources seem to be galactic.

\noindent
$\bullet\ $Only a small fraction of them will be pulsars and SNR.

\noindent
$\bullet\ $The majority of them apparently belongs to other, new classes of gamma-ray emitters.

\noindent
$\bullet\ $Diffuse gamma-ray excesses have been observed at energies of
around 100 keV and at a few GeV.

\noindent
$\bullet\ $The GeV excess is probably not caused by unresolved point sources.

\noindent
$\bullet\ $Hard IC models of the GeV excess are not in a serious conflict with
the non-detection of SNR or SNR halos by EGRET.

\section{Acknowledgements}
Partial support by the Bundesministerium f\"ur Bildung und Forschung through
DESY, grant 05CH1PCA/6, is gratefully acknowledged.  

\vspace{1pc}
\re
Aharonian F.A.\ et al.\ 2002, A\&A 393, L37
\re
Aharonian F.A.\ et al.\ 2001, A\&A 375, 1008
\re
Berezhko E.G. \& V\"olk H.J.\ 2000, ApJ 540, 923
\re
Cheng K.S. \& Zhang L.\ 1998, ApJ 498, 327
\re
Enomoto R.\ et al.\ 2002, Nature 416, 823
\re
Gehrels N.\ et al.\ 2000, Nature 404, 363
\re
Grenier I.A.\ 2000, A\&A 364, L93
\re
Harding A.K. \& Zhang B.\ 2001, ApJ 548, L37
\re
Hartman R.C.\ et al. 1999, ApJS 123, 79
\re 
Hunter S.D.\ et al.\ 1997, ApJ 481, 205
\re
Kaaret P. \& Cottam J.\ 1996, ApJ 462, L35
\re
Kinzer R.L., Purcell W.R., Kurfess J.D.\ 1999, ApJ 515, 215
\re
LeBohec S.\ et al.\ 2000, ApJ 539, 209
\re 
McLaughlin M.A. \& Cordes J.M.\ 2000, ApJ 538, 818
\re
McLaughlin M.A., Mattox J.R., Cordes J.M., Thompson D.J.\ 1996, ApJ 473, 763
\re
Merck M.\ et al.\ 1996, A\&AS 120, C465
\re
Pohl M.\ 1998, A\&A 339, 587 
\re
Pohl M., Esposito J.A.\ 1998, ApJ 507. 327
\re
Pohl M.\ 1997, ApJ 491, 164
\re 
Purcell W.R.\ et al.\ 1996, A\&AS 120, C389
\re 
Reimer O. \& Pohl M.\ 2002, A\&A 390, L43
\re
Reynolds S.P. \& Keohane J.W.\ 1999, ApJ 525, 368 
\re
Romero G.E., Benaglia P., Torres D.F.\ 1999, A\&A 348, 868
\re
Skibo J.G., Ramaty R., Purcell W.R.\ 1996, A\&AS 120, C403
\re
Sreekumar P.\ et al.\ 1998, ApJ 494, 523
\re
Strong A.W., Moskalenko I.V., Reimer O.\ 2000, ApJ 537, 763
\re 
Strong A.W.\ et al.\ 1996, A\&AS 120, C381
\re 
Sturner S.J. \& Dermer C.D.\ 1995, A\&A 293, L17
\re
Torres D.F.\ et al.\ 2001, A\&A 370, 468
\re 
Valinia A., Kinzer R.L., Marshall F.E.\ 2000, ApJ 534, 277
\re 
Valinia A. \& Marshall F.E.\ 1998, ApJ 505, 134
\re
Yadigaroglu I.-A. \& Romani R.W.\ 1995, ApJ 449, 211
\re
Zhang L. \& Cheng K.S.\ 1998, A\&A 335, 234

\chapter*{ Entry Form for the Proceedings }

\section{Title of the Paper}

{\Large\bf %
Diffuse emission from the Galactic plane and unidentified EGRET sources
}

\section{Author(s)}

I am the author of this paper.

\newcounter{author}
\begin{list}%
{Author No. \arabic{author}}{\usecounter{author}}

\item %
\begin{itemize}
\item Full Name:                Martin Pohl 
\item First Name:               Martin 
\item Middle Name:               
\item Surname:                  Pohl 
\item Initialized Name:         M. Pohl 
\item Affiliation:              Ruhr-Universit\"at Bochum, Germany 
\item E-Mail:                   mkp@tp4.ruhr-uni-bochum.de 
\item Ship the Proceedings to:  Institut f\"ur Theoretische Physik, Lehrstuhl 4,
Ruhr-Universit\"at Bochum, Germany 
\end{itemize}

\end{list}

\endofpaper
\end{document}